\documentclass[pra,aps,groupedaddress,twocolumn,floatfix,nofootinbib]{revtex4}
\usepackage{graphicx}

\newcommand{\beq}{\begin{equation}}
\newcommand{\eeq}{\end{equation}}
\newcommand{\beqa}{\begin{eqnarray}}
\newcommand{\eeqa}{\end{eqnarray}}

\newcommand{\braket}[2]{\mbox{$ \langle #1 | #2 \rangle $}}
\newcommand{\ketbra}[2]{\mbox{$ | #1 \rangle\!\langle #2 |$}}

\newcommand{\ket}[1]{\mbox{$ | #1 \rangle $}}
\newcommand{\bra}[1]{\mbox{$ \langle #1 | $}}

\def\up{\uparrow}
\def\down{\downarrow}

\def\half{\frac{1}{2}}
\def\opone{\leavevmode\hbox{\small1\normalsize\kern-.33em1}}
\def\qe{\stackrel{?}{=}}

\begin{document}

\title{Quantum Measurements, Energy Conservation and Quantum Clocks}
%===================================================================
\author{Nicolas Gisin and Emmanuel Zambrini Cruzeiro}
\affiliation{
    Group of Applied Physics, University of Geneva, 1211 Geneva 4, Switzerland}

\date{\small \today}
\begin{abstract}
We consider a spin chain extending from Alice to Bob with next neighbors interactions, initially in its ground state. Assuming that Bob measures the last spin of the chain, the energy of the spin chain has to increase, at least on average, due to the measurement disturbance. Presumably, the energy is provided by Bob's measurement apparatus. Assuming now that, simultaneously to Bob's measurement, Alice measures the first spin, we show that either energy is not conserved, - implausible - or the projection postulate doesn't apply, and that there is signalling. An explicit measurement model shows that energy is conserved (as expected), but that the spin chain energy increase is not provided by the measurement apparatus(es), that the projection postulate is not always valid - illustrating the Wigner-Araki-Yanase (WAY) theorem - and that there is signalling, indeed. The signalling is due to the non-local interaction Hamiltonian. This raises the question of a suitable quantum information inspired model of such non-local Hamiltonians.
\end{abstract}
\maketitle

\section{Introduction}\label{intro}
%=====================
Consider a quantum system that extends over a distance from Alice to Bob, for instance a spin chain. Assume that the system is initially in its ground state and that this is an entangled state. Alice and Bob have the choice between either doing nothing or performing a predetermined measurement. If they chose to measure their part of the system, the system gets perturbed, hence does not remain in the ground state. Accordingly, its mean energy increases, borrowing - presumably - energy from the measurement apparatus\footnote{We include the local environment in the measurement apparatus.} (or apparatuses if both Alice and Bob perform a measurement). This scenario illustrates several interesting questions. Can Bob determine whether Alice did measure her part of the system by merely looking at the energy change of his measurement apparatus? Is there signalling and what does signalling mean in this non-relativistic scenario? Measuring an energy change requires time \cite{AMPTime}, hence there is a trade-off between the energy measurement accuracy and the required measurement time. Finally, how should one model energy measurements in a finite time?

We analyse this simple situation and see that there is signalling, indeed, but nothing non-physical. We also learn that, strictly speaking, it is not the measurement apparatuses that provide the energy. Finally, we see that the standard projection postulate can't be blindly applied to sharp energy measurements. We also recall how one can model realistic energy measurements using quantum clocks. 

Somehow, all this can be found in a diversity of literature \cite{LiebRobinson,EisertLR,VerstreateLiebRobinson,WAYWigner,WAYAraki,WAYYanase,WAYBusch,BuschQTM}, but it seems poorly known and is usually not presented and illustrated in a concise form. We shall consider spin chains, a subject usually studied by solid state theorists, but our approach is in the spirit of quantum information. For example, in solid state next neighbors interaction Hamiltonians are called local because their range extends only over a finite distance, while the same Hamiltonians are thought of as nonlocal in quantum information because they allow one to entangled distant spins. Solid state theorists know well the Lieb-Robinson light-cone-like bound on information propagation in spin chains \cite{LiebRobinson,EisertLR,VerstreateLiebRobinson}, but are not too concerned about the exponentially small information propagating outside that cone, while quantum information is much concerned about any superluminal information. Also, the WAY theorem \cite{WAYWigner,WAYAraki,WAYYanase,WAYBusch} is mostly known in solid state, but possibly not in the form we illustrate in this paper. Finally, several intriguing questions remain open.

\section{The 3-spin chain example}\label{threeSpins}
\begin{figure}[h]
\includegraphics[width=8.5cm]{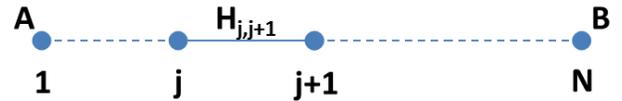}
\vspace{-10pt}
\caption{\it A linear chain of spin $\half$ with next neighbors interactions. Alice holds the first spin and Bob the last one.}
\label{FigSpinChain}
\end{figure}

%===================================
Consider a chain of spin $\half$ with next neighbors interactions, see Fig. \ref{FigSpinChain}. For simplicity we limit ourselves to a 3-spin chain, though longer chains would equally work, and, admittedly, a mere 2 spin system would also suffice. But let's consider 3 spins on a line, Alice holding the first spin and Bob the third spin, with the following next neighbors interaction:
\beq\label{Hint}
H_{j,j+1}=\frac{1}{2\sqrt{2}}(\sigma_x^j\otimes\sigma_x^{j+1}-\sigma_y^j\otimes\sigma_y^{j+1})
\eeq
The coefficient is chosen to simplify the eigenvalues.

Note that this Hamiltonian can equivalently be written in terms of the Bell states $\Phi^\pm=\frac{1}{\sqrt{2}}(\ket{\up\up}\pm\ket{\down\down})$: $H_{j,j+1}=\frac{1}{\sqrt{2}}(P_{\Phi^+}-P_{\Phi^-})$, where $P_{\Phi^\pm}\equiv\ketbra{\Phi^\pm}{\Phi^\pm}$ denotes the projector on state $\ket{\Phi^\pm}$. Hence, in the ground state each neighbor pair of spins should be close to the $\Phi^-$ state.

The eigenstates of the 3-spin Hamiltonian
\beq
H\equiv H_{1,2}+H_{2,3}
\eeq
read as follows, where the lower index indicates the corresponding eigenvalue ($H\ket{\phi_E}=E\ket{\phi_E}$):
\beqa
\phi_{-1}&=&\frac{1}{\sqrt{2}}\ket{\up\up\up}-\frac{1}{2}\ket{\up\down\down+\down\down\up}\label{phim1}\\
\phi_0^1&=&\ket{\up\down\up}\\
\phi_0^2&=&\frac{1}{\sqrt{2}}\ket{\up\down\down-\down\down\up}\\
\phi_{+1}&=&\frac{1}{\sqrt{2}}\ket{\up\up\up}+\frac{1}{2}\ket{\up\down\down+\down\down\up}
\eeqa
By symmetry of the Hamiltonian, flipping all 3 spins doesn't change the energy eigenvalue, hence the 4 other eigenstates are obtained by flipping all 3 spins. Accordingly, there are 3 eigenvalues: $\pm1$ doubly degenerated and 0 4-times degenerated.

We assume the 3-spin system is initially in state $\phi_{-1}$, hence the energy of the system is initially minimal: $E_{in}=-1$.

\section{Case where only Bob measures $\sigma_z$}\label{Bmeasures}
%================================================
Let's first assume that only Bob performs a measurement, $\sigma_z$, while Alice remains passive.
Denote $b$ Bob's measurement result and $\overline{\Delta E_b}$ the mean energy increase of the 3-spin chain in case Bob got the result $b=\pm1$.

Consider the outcome $b=+1$ which happens with probability $\frac{3}{4}$ (see eq. \ref{phim1}). Conditioned on this $b=+1$ outcome, the energy increases on average by 1/3 unit:
\beqa
\overline{\Delta E_{b=+1}}&=&\frac{\bra{P_\up^B\phi_{-1}}H\ket{P_\up^B\phi_{-1}}}{\bra{\phi_{-1}}P_\up^B\ket{\phi_{-1}}}-E_{in}\\
&=&\frac{-2}{3}-(-1)=\frac{1}{3}
\eeqa
where $P_\up^B$ denotes the projector on Bob's spin up.

Note that if the energy of the 3-spin chain is measured sharply, the energy increase is found to be either 0, 1 or 2 units - as the energy eigenstates are 0 and $\pm1$ - with probabilities $\frac{3}{4}$, $\frac{1}{6}$ and $\frac{1}{12}$, respectively. The average recovers 1/3, as it should. Finally, the mean energy increase, averaged over all possible results $b=\pm1$, equals 1/2. All cases are summarized in Table \ref{table1} of Appendix \ref{appendixA}.

In summary, let's remember that for outcome $b=+1$, the energy increase is $+1/3$.

This energy increase has to come from somewhere. Since the 3-spin chain interacted only with Bob's measurement apparatus, it has to be the case that the energy was provided by this measurement apparatus and Bob should be able to measure his apparatus and find that it has lost energy $-1/3$ on average. 

Note that since the situation is symmetric between Alice and Bob, the 3-spin system undergoes the same energy increase of $+1/3$ in the case where only Alice performs a $\sigma_z$ measurement and gets $a=+1$.

\section{Case where both Alice and Bob measure $\sigma_z$}\label{ABmeasure}
%========================================================
Next, let's assume that Alice and Bob measure $\sigma_z$ on their spins with results $a$ and $b$. Assume for simplicity that both measurements are simultaneous. 
Consider the case where both Alice and Bob get result $+1$: $a=b=+1$. This happens with probability $\half$. In this case the averaged energy increase is $+1$:
\beqa
\overline{\Delta E_{a=b=+1}}&=&\frac{\bra{P_{\up\up}^{AB}\phi_{-1}}H\ket{P_{\up\up}^{AB}\phi_{-1}}}{\bra{\phi_{-1}}P_{\up\up}^{AB}\ket{\phi_{-1}}}-E_{in}\\
&=&0-(-1)=1
\eeqa
where $P_{\up\up}^{AB}=P_\up^A\otimes\opone\otimes P_\up^B$ denotes the tensor product projector on Alice and Bob's spins up.

This leads to a conundrum: since Alice and Bob are at a distance and perform their measurements simultaneously, Alice and Bob's measurement apparatuses each provide the average energy $+1/3$, as if the other did not perform any measurement. But then, where does the third 1/3 energy increase come from? 

A second aspect of this conundrum goes as follows. By symmetry, if both Alice and Bob measure simultaneously their spins, then the energy increase $+1$ should be provided equally by both measurement apparatuses, $+1/2$ each. But then, Bob could detect whether Alice did perform a measurement or not, just by looking at the energy change of his apparatus. Does that lead to (unphysical) signalling \cite{GisinQchance14}?

Several observations are due here:

First, we conditioned all the above analysis on Alice and/or Bob, whoever measured, getting result $+1$. Hence, Bob would have to wait until he knows Alice's result, an information he can't get faster than light. Consequently, this is not signalling in the sense of faster than light communication. Nevertheless, it seems obvious that the energy of Bob's measurement apparatus doesn't change when he receives the information about Alice's result. Hence, the conundrums remain.

Second, in quantum theory signalling has nothing to do with the speed of light $c$: there is no $c$ in any Hilbert space. Signalling is unphysical communication, i.e. information transfer without any physical system carrying this information, as in Newtonian gravity \cite{GisinQchance14}.

Third, it is well-known that the Lieb-Robinson bound describes the propagation of information in any spin chain (and lattice) as bounded by some finite velocity \cite{LiebRobinson,EisertLR,VerstreateLiebRobinson}. Only an exponentially small amount of information propagates outside this light-cone-like cone. Here, however, we are concerned with this small amount of information that seems to propagate at an arbitrarily fast speed.

Actually, a more involved analysis of all possible cases reveals that signalling remains also if one doesn't condition on the result $+1$. But for the fluidity of our story, we relegate this more involved analysis to Appendix \ref{appendixA}.

\section{What could go wrong?}\label{wrong}
%=============================
So, what could have gone wrong? We merely assumed energy conservation and the standard projection postulate.

Some readers may have heard that additive conserved quantities, like e.g. energy, can't be measured accurately in a reproducible way, e.g. that the projection postulate is inadequate for energy measurements \cite{WAYWigner,WAYAraki,WAYYanase,WAYBusch}. Moreover, many readers are aware that energy measurements generally take time\footnote{As proven by Aharonov and Bohm long ago, this is not necessarily the case provided the Hamiltonian is known \cite{AharonovBohmTime}. However, as shown more recently by Aharonov, Massar and Popescu, in cases where the Hamiltonian is not known, there is a tradeoff between the measurement time and energy accuracy, indeed \cite{AMPTime}. In any case, here the 3-spin Hamiltonian is known, but Alice and/or Bob measure only one of the spins.}. But the consequence for our story of this last point is unclear. Moreover, the length of the spin chain could be increased with similar conclusions; hence the Alice-Bob distance could seemingly be adapted to whatever finite duration energy measurements require.

In order to get a clear understanding of the situation we need to put the measurement apparatus(es) within a larger autonomous quantum system. Then energy conservation follows directly from the Schr\"odinger equation, isn't it? The standard model of quantum measurements, due to Von Neumann \cite{QMVN}, uses time-dependent Hamiltonians, as we recall in the next section. But energy conservation with time-dependent Hamiltonians is unclear. Hence, we need to incorporate in a grand autonomous quantum system also clocks, actually quantum clocks. This is done in section \ref{meas3qubits}. To prepare the ground for this we introduce in the next section energy measurement models of a single spin, first in the energy eigenbasis (sub-section \ref{energyBasis}), next in an arbitrary basis (sub-section \ref{nonEnergyBasis}) and finally two simultaneous measurements (sub-section \ref{simulMeas}).

\section{Models of quantum measurements}\label{meas}
%=======================================
The standard model of quantum measurements goes back to Von Neumann \cite{QMVN}. Consider an observable $A$ and a (1-dimensional) pointer with position and momentum operators $q$ and $p$, respectively. The system and the pointer are coupled by the interaction Hamiltonian $H_{int}(t)=g(t) A\otimes p$, where $g(t)$ is a function of time non-vanishing only for a short time close to the time of the measurement, $t_m$, and such that $\int g(t)dt=1$. During the interaction $g(t)$ is assumed large enough that the interaction Hamiltonian dominates all other Hamiltonians. Denote $\Phi_0$ the initial state of the pointer, e.g. a Gaussian centered around position 0. Denoting $\psi_0$ the initial state of the system, one has:
\beqa
e^{-i\int H_{int}(t)dt}\psi_0\otimes\Phi_0&=&e^{-iA\otimes p}\psi_0\otimes\Phi_0 \nonumber\\
&=&\sum_a \braket{a}{\psi_0}~\ket{a}\otimes\Phi_a
\eeqa
where the $a$'s and $\ket{a}$'s denote the eigenvalues and eigenvectors of the observable $A$, respectively, and $\Phi_a\equiv e^{-ia\cdot p}\Phi_0$ is the state of the pointer displaced by a distance corresponding to the eigenvalue $a$. If the pointer's position quantum indeterminacy (often called uncertainty) is smaller than the distance between the eigenvalues of $A$, then a position measurement of the pointer determines the measurement result of the observable $A$: $x_{pointer}\approx a$. Such measurements are called strong or sharp. The other cases, when the indeterminacy is not negligible, are called weak. 

Let us stress explicitly that all strengths of measurements can thus be modeled - not only ultra-weak measurements - and that this model has nothing to do with the notion of weak values as developed by Aharonov and coworkers\footnote{This comment is mostly for standard referees. Indeed, in all my previous papers using weak measurements there has always been at least one referee who confused weak measurements with weak values.} \cite{weakValue}. 

In this paper we consider only strong measurements with binary outcomes. Consequently, it suffices to consider pointers living in a 2-dimensional Hilbert space, i.e. the pointer is a mere qubit. Note that this qubit could represent a large quantum system, like a pointer with basis states $\ket{point~ to~ label~ "spin~ up"}$ and  $\ket{point~ to~ label~ "spin~ down"}$.

Accordingly, a measurement of $A=\sigma_z$ can be modeled with the following interaction Hamiltonian: 
\beq
H_{int}=\frac{\pi}{2}g(t) P_\up^{spin}\otimes\sigma_x^{pointer}
\eeq
where $P_\up^{spin}\equiv\ketbra{\up}{\up}$ denote the projector on the spin-up state.
The spin and qubit-pointer evolve:
\beqa
&& e^{-i\int H_{int}(t)dt}\psi_0\otimes\ket{0} \nonumber\\
&&=e^{-i\frac{\pi}{2} P_\up^{spin}\otimes\sigma_x^{pointer}}\psi_0\otimes\ket{0} \\
&&=\braket{\up}{\psi_0}\ket{\up}_{spin}\otimes\ket{1}_{pointer}+\braket{\down}{\psi_0}\ket{\down}_{spin}\otimes\ket{0}_{pointer}\nonumber
\eeqa

The Von Neumann model is a useful tool. However, since the interaction Hamiltonian depends on time, energy is not conserved. Hence this model is not suitable to analyze our spin chain example. In addition to the system and the pointer we need a quantum clock that determines when the system-pointer interaction is on \cite{AharonovQrefframe,AharonovArrivalTime}. As clock we consider a simple 1-dimensional wave packet (possibly with bounded support) moving along a line with the Hamiltonian $p$. This Hamiltonian is unbounded from below, hence unphysical, but sufficient for our purpose. Note that the clock's velocity is unity, hence time can be directly read from the clock's position. The spread of the clock's wave-packet remains constant and corresponds to the time uncertainty.

In order to gain familiarity with this measurement model, the following sub-sections considers measurements on a single spin $\half$.

\subsection{Measurements of a single spin in the energy eigenbasis}\label{energyBasis}
\begin{figure}
\includegraphics[width=9cm]{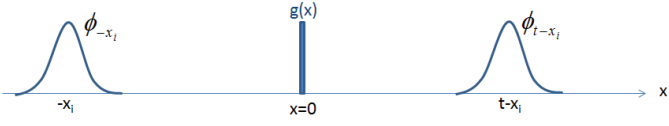}
\vspace{-10pt}
\caption{\it The quantum clock starts on the far left, around position $-x_i$, triggers the spin-pointer interaction near position 0, and goes on to the right around position $t-x_i$.}
\label{FigQclock}
\end{figure}

%==================================================================
Consider a system consisting of the following 3 subsystems: a spin $\half$, a qubit (the pointer) and a 1-dimensional clock. The initial state of the qubit-pointer is $\ket{0}$ and the clock starts in a state $\phi_{-x_i}$ of width $\Delta$ centered around $-x_i$ on the far left of the 1-dimensional axis, see Fig. \ref{FigQclock}. The interaction takes place when the clock is around the origin of that axis. We assume that the pointer has no free Hamiltonian, while the spin's Hamiltonian is $\omega\sigma_z$.

In this sub-section we consider the case where the spin is measured along the z-direction, i.e. the observable is $\sigma_z$. The full (time independent) Hamiltonian reads thus:
\beq
H=\omega\sigma_z^{spin}+p^{clock}+\ketbra{\up}{\up}\otimes\sigma_x^{pointer}\otimes g(q^{clock})
\eeq
where the function $g$ is nonzero only in a small interval near the origin, see Fig. \ref{FigQclock}; its norm is such that when the clock is near the origin and if the spin points ``up'', then the qubit-pointer gets flipped (action by $\sigma_x^{pointer}$ with a coupling $\int_{-\infty}^\infty g(x)dx=\pi/2$).

\begin{figure}[h]
\includegraphics[width=7cm]{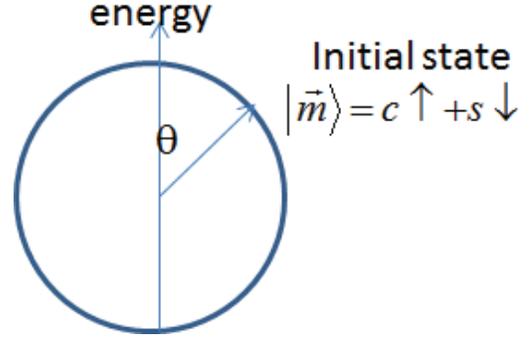}
\vspace{-10pt}
\caption{\it Sketch of the configuration of a quantum measurement of the energy of a spin.}
\label{FigEnergyMeas}
\end{figure}

If the spin's initial state is $\ket{\vec m}=c\ket{\up}+s\ket{\down}$, see Fig. \ref{FigEnergyMeas}, then, for $t>>x_i$:
\beq
e^{-iHt}\ket{\vec m,0,\phi_{-x_i}}=\big(ce^{-i\omega t}\ket{\up,1}+se^{i\omega t}\ket{\down,0}\big)\ket{\phi_{t-x_i}}
\eeq
where we dropped the suffixes: we always write the quantum states in the order spin-pointer-clock.

Accordingly, the pointer's state is fully correlated with the spin energy eigenstate and this remain so for all times $t>>x_i$. Consequently, reading the pointer provides full information about the spin's state and energy. This illustrates that when the (spin) Hamiltonian is known, the energy can be measured in an arbitrarily short time \cite{AharonovBohmTime}. But this is not true when the Hamiltonian is unknown \cite{AMPTime}.

As a side remark, we note that if such a measurement model is applied to a photon number measurement, $A=a^\dagger a$, on a coherent photonic state, the pointer (which would need to have an countably infinite number of orthogonal states) merely reveals the number of photons in the initial coherent state. This agrees well with the fact that when there is no phase reference, as in photon number measurements, a coherent state is indistinguishable from a Poisson photon-number statistical distribution for which a photon number measurement merely reveals the pre-existing number of photons.

\subsection{Measurement of a single spin in a basis incompatible with the energy: it is not the pointer that provides the energy}\label{nonEnergyBasis}
%================================================================================

\begin{figure}[h]
\includegraphics[width=7cm]{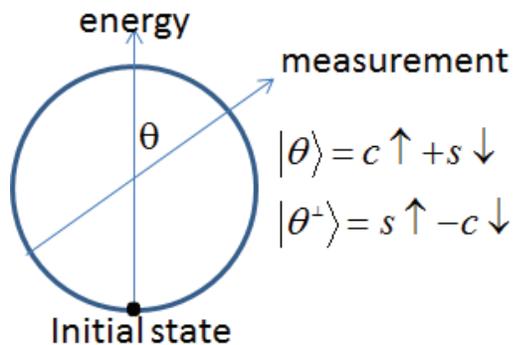}
\vspace{-10pt}
\caption{\it Sketch of the configuration of a quantum measurement on a spin in a basis incompatible with the energy.}
\label{FigQmeas}
\end{figure}

Here we apply our measurement model to the more interesting case in which the spin Hamiltonian $\omega\sigma_z$ does not commute with the measured observable $\ketbra{\theta}{\theta}$, where $\ket{\theta}=c\ket{\up}+s\ket{\down}$, see Fig. \ref{FigQmeas}:
\beq
H=\omega\sigma_z^{spin}+p^{clock}+\ketbra{\theta}{\theta}\otimes\sigma_x^{pointer}\otimes g(q^{clock})
\eeq
For simplicity assume that the spin is initially ``down''.

We start by assuming that the clock is in a delta function, i.e. $\Delta=0$. Once this case solved, the full solution can be obtained by linearity of the Schr\"odinger equation.
\beqa
\ket{\down,0,-x_i}&\stackrel{t=x_i-\epsilon}{\rightarrow}&\big(s\ket{\theta,0}-c\ket{\theta^\perp,0}\big)\ket{-\epsilon} \nonumber\\
&\stackrel{t=x_i+\epsilon}{\rightarrow}&\big(s\ket{\theta,1}-c\ket{\theta^\perp,0}\big)\ket{+\epsilon} \nonumber\\
=\big(sc&\ket{\up,1}+&s^2\ket{\down,1}-sc\ket{\up,0}+c^2\ket{\down,0}\big)\ket{+\epsilon} \nonumber\\
&\stackrel{t=x+x_i}{\rightarrow}&\big(  sc\cdot e^{-2i\omega x}\ket{\up,1} +s^2\ket{\down,1} \label{diagMeas}\\ 
&&  -sc\cdot e^{-2i\omega x}\ket{\up,0} +c^2\ket{\down,0} \big)\ket{x}  \nonumber
\eeqa
Note that the phases in the above expression are very important. Indeed, for all states $\ket{\phi}$ of the clock, one has:
\beq
\bra{e^{-2i\omega x}\phi}p\ket{e^{-2i\omega x}\phi}=\bra{\phi}p\ket{\phi}-2\omega
\eeq
Consequently, the first and third terms in (\ref{diagMeas}) indicate that the spin was flipped by the measurement and that the increase of energy was provided by the clock. As anticipated in the introduction, it is thus not the measurement apparatus, stricto sensu the pointer, that provides the energy. Note that this had to be expected: since the pointer carries no energy, any energy change of the spin has to be compensated by energy change in the clock. It should be stressed that the change in the expectation value of the ``momentum'' of the pointer doesn't mean that the pointer changes speed (with respect to the outside time). Indeed, recall that we use the (unphysical) clock Hamiltonian $p$, hence the operator $p$ should not be thought of as ``momentum'', but rather as a mere displacement operator. Hence, shifting $p$ affects the energy, but not the velocity of the clock (only the global phase may be affected), nor its spread that remains unchanged.

By superposing many delta functions one can recover all possible initial wave-functions. Denote \ket{(x,p)} a wave-packet with mean position and momentum $x$ and $p$, respectively\footnote{For simplicity we assume Gaussian-like wave-packets and we don't consider wave-packets with complex structures, e.g. with several peaks.}. One has, for $t>>x$:
\beqa
\ket{\down,0,(-x,p)}&\stackrel{t}{\rightarrow}&sc\ket{\up,0,(t-x,p-2\omega)} \nonumber\\
&+&s^2\ket{\down,0,(t-x,p)} \nonumber\\
&-&sc\ket{\up,1,(t-x,p-2\omega)} \nonumber\\
&+&c^2\ket{\down,1,(t-x,p)} \label{phit}
\eeqa
up to an irrelevant global phase $e^{i\omega t}$. Note that in the 1st and 3rd components above the spin is flipped, hence increases its energy by $2\omega$, which is precisely compensated by the clock's energy shifts of $-2\omega$.

The solution (\ref{phit}) is interesting. Assume first that the pointer's momentum indeterminacy - and hence the pointer's energy indeterminacy - is large, i.e. $\ket{(x,p)}\approx\ket{(x,p-2\omega)}$. In this case (\ref{phit}) simplifies to:
\beq
\ket{\down,0,(-x,p)}\stackrel{t}{\rightarrow}\big(s\ket{\theta,0}-c\ket{\theta^\perp,1}\big)\ket{(t-x,p)} 
\eeq
which corresponds to merely applying the usual projection postulate to the initial state $\ket{\vec m}$: if the pointer's state reads $\ket{0}$ or $\ket{1}$, then the spin state is $P_\theta\ket{\down}$ or $P_{\theta^\perp}\ket{\down}$, respectively. In this case energy conservation is badly defined, as the initial and final energies are poorly determined. Note that the clock factorizes.

Next, consider the case where the clock's momentum is sharply determined, hence the clock's energy is sharply determined. In this case, measuring the pointer does not lead to a pure state of the spin, but in a spin-clock entangled state, such that the total energy is well determined and conserved. Furthermore, if one merely traces out (i.e. ignores) the clock, then the spin ends up in a mixed state:
\beq
\ket{0}_{pointer}\rightarrow\rho_{spin}=s^2c^2\ketbra{\up}{\up}+s^4\ketbra{\down}{\down}
\eeq
and similarly for the other possible result. Note that $Tr(\rho_{spin})=prob(result~\ket{\theta})=s^2=|\braket{\theta}{\down}|^2$; the Born rule thus holds, but the post-measurement state differs from the one predicted by the projection postulate.

Let's emphasize that in this case of well determined energy one can't merely apply the projection postulate: in this case the projection postulate is not valid, as anticipated long ago by Wigner, Araki and Yanase, i.e. the WAY theorem \cite{WAYWigner,WAYAraki,WAYYanase}, for a recent review see \cite{WAYBusch}. Let's recall that the WAY theorem states that additive conserved quantities can't be measured in a reproducible way. Since energy is a conserved additive quantity and since reproducible measurements, sometimes called ideal measurements, are described by the projection postulate, our result illustrates the WAY theorem\footnote{As careful readers will have noticed, in this subsection we do not measure energy, but measure a spin component in an arbitrary direction $\theta$. This corresponds to the case where the Hamiltonian is unknown. In such a case, all one can do is to perform some measurements, possibly weak measurements, in an arbitrary direction as in this subsection.}.

\subsection{Simultaneous measurements on one spin}\label{simulMeas}
\begin{figure}[h]
\includegraphics[width=8.5cm]{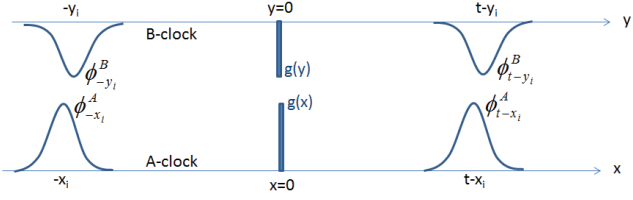}
\vspace{-10pt}
\caption{\it Two quantum clocks that trigger, when around position 0, quantum measurements on a single spin. The initial states of the two clocks, $\phi_{x_i}^A$ and $\phi_{y_i}^B$, determine whether the two measurements are successive, simultaneous or partially overlapping. Note that the two clocks could be initially entangled.}
\label{FigSimulQmeas}
\end{figure}

%==================================================
In order to get ready to apply our measurement model to our 3-spin example, let's now model a simultaneous measurement of a single spin with two pointers, the A-pointer and B-pointer, each associated to a clock, the A-clock and B-clock, respectively, see Fig. \ref{FigSimulQmeas}. For simplicity we assume that the two simultaneous measurements are compatible (even identical for this simple example), but not in the energy basis:
\beqa
H&=&\omega\sigma_z+p_x^{Aclock}+p_y^{Bclock} \nonumber\\
&+&\ketbra{\theta}{\theta}\otimes\sigma_x^{Apointer}\otimes g(q^{Aclock}) \nonumber\\
&+&\ketbra{\theta}{\theta}\otimes\sigma_x^{Bpointer}\otimes g(q^{Bclock}) 
\eeqa

A computation very similar to that of the previous sub-section leads to a result similar to (\ref{diagMeas}):
\beqa \label{PhitAB}
\phi_t^{AB}(x,y)&=&\big\{sc\cdot e^{-2i\omega\cdot\max(x,y)}\ket{\up,1-0,c^2\cdot1+s^2\cdot0} \nonumber\\
&+&sc\cdot e^{-2i\omega\cdot\min(x,y)}\ket{\up,s^2\cdot1+c^2\cdot0,1-0} \nonumber\\
&+&s^2c^2\cdot e^{-2i\omega|x-y|}\ket{\down,1-0,1-0} \nonumber\\
&+&\ket{\down,s^2\cdot1+c^2\cdot0,s^2\cdot1+c^2\cdot0}\big\} \nonumber\\
&\cdot&\Phi_t(x,y)
\eeqa
where $\Phi_t(x,y)=\Phi_0(x-t,y-t)$ and $\ket{s^2\cdot1+c^2\cdot0}\equiv s^2\ket{1}+c^2\ket{0}$. Note that $e^{-2i\omega|x-y|}=e^{-2i\omega(\max(x,y)-\min(x,y))}$. The presence of $\max(x,y)$ and $\min(x,y)$ in eq. (\ref{PhitAB}) indicates that it matters which pointer interacts first with the spin. To illustrate this, assume $x_i<<y_i$, i.e. the A-pointer measures the spin before the B-pointer. In this case, using notation as in (\ref{phit}), eq. (\ref{PhitAB}) simplifies to:
\beqa\label{PhitAaB}
\phi_t^{A<B}=\hspace{7cm}&&\\
sc\ket{\up,1-0,c^2\cdot1+s^2\cdot0}\ket{(t\!-\!x_i,p_A\!-\!2\omega),(t\!-\!y_i,p_B)}&&\nonumber\\
+sc\ket{\up,s^2\cdot1+c^2\cdot0,1-0}\ket{(t-x_i,p_A),(t\!-\!y_i,p_B\!-\!2\omega)}&&\nonumber\\
+\ket{\down,s^2\cdot1+c^2\cdot0,s^2\cdot1+c^2\cdot0}\ket{(t\!-\!x_i,p_A),(t\!-\!y_i,p_B)}&&\nonumber\\
+s^2c^2\ket{\down,1-0,1-0}\ket{(t\!-\!x_i,p_A\!-\!2\omega),(t\!-\!y_i,p_B\!+\!2\omega)}&&\nonumber
\eeqa
The first line in the equation above represents the case where the A-clock provides the energy for the spin flip, the second line the case where the B-clock provides the energy, the third line the case where no energy is exchanged and the fourth line the case where Alice's clock provides energy and Bob's clock takes it away, i.e. the case where there are 2 spin flips. This clearly entangles the clocks and the spin, even after the pointers are measured.

In Appendix \ref{appendixB} we analyse the state of the spin and clocks once the pointers have been measured. This illustrates again that the projection postulate is not always valid \cite{BuschQTM}.

The above result calls for an analysis when the two clocks are initially entangled. However, since this is not the main purpose of this paper we defer this analysis to Appendix \ref{appendixC}.

\section{Measurements on our 3-spin examples with quantum clocks}\label{meas3qubits}
%=================================================================
We are now in position to apply our measurement model to our 3-spin chain example. The global system, consisting of the 3-spin chain and of Alice and Bob's pointers and clocks, undergoes an autonomous evolution governed by the time-independent Hamiltonian:
\beqa
H&=&H_{1,2}+H_{2,3}+p_x^{Aclock}+p_y^{Bclock} \nonumber\\
&+&\ket{\up}\!_A\!\bra{\up}\otimes\sigma_x^{Apointer}\otimes g(q^{Aclock}) \nonumber\\
&+&\ket{\up}\!_B\!\bra{\up}\otimes\sigma_x^{Bpointer}\otimes g(q^{Bclock}) 
\eeqa
where the interaction Hamiltonians $H_{j,j+1}$ are given by (\ref{Hint}). Note that the measurements are compatible, i.e. $\ket{\up}\!_A\!\bra{\up}$ and $\ket{\up}\!_B\!\bra{\up}$ commute, but are not compatible with the 3-spin energy. The computation is similar to those of the previous section. The full solution, analogous to (\ref{PhitAB}), reads:
\beqa\label{Phit3AB}
&&\ket{\phi_{-1},0_A,0_B,\Phi_0(x,y)}\stackrel{t}{\rightarrow} \phi_t^{AB} \nonumber\\
&=&\big\{\ket{\phi_{-1},\frac{3\cdot0+1}{4},\frac{3\cdot0+1}{4}} \nonumber\\
&+&\ket{\phi_{+1},\frac{3\cdot0+1}{4},\frac{0-1}{4}}e^{-4i\omega\cdot\min(x,y)} \nonumber\\
&+&\ket{\phi_{0},\frac{3\cdot0+1}{4},\frac{0-1}{\sqrt{8}}}e^{-2i\omega\cdot\min(x,y)} \nonumber\\
&+&\ket{\phi_{+1},\frac{0-1}{4},\frac{3\cdot0+1}{4}}e^{-4i\omega\cdot\max(x,y)} \nonumber\\
&+&\ket{\phi_{-1},\frac{0-1}{4},\frac{0+-1}{4}}e^{-4i\omega\cdot|x-y|} \nonumber\\
&-&\ket{\phi_{0},\frac{0-1}{4},\frac{0-1}{\sqrt{8}}}e^{-2i\omega\cdot(2\max(x,y)-\min(x,y))} \nonumber\\
&+&\ket{\phi_{+1},\frac{0-1}{\sqrt{8}},\frac{0-1}{\sqrt{8}}}e^{-2i\omega\cdot(\max(x,y)+\min(x,y))} \nonumber\\
&-&\ket{\phi_{-1},\frac{0-1}{\sqrt{8}},\frac{0-1}{\sqrt{8}}}e^{-2i\omega\cdot|x-y|} \nonumber\\
&-&\ket{\phi_{0},\frac{0-1}{\sqrt{8}},\frac{0+1}{\sqrt{8}}}e^{-2i\omega\cdot\max(x,y)}\big\}\phi_0(x-t,y-t)
\eeqa
Everything can be computed from this last expression. For example, Appendix \ref{appendixD} presents the case where Alice measures before Bob, with notations analog to (\ref{PhitAaB}), illustrating all possible energy exchanges that may happen.

The general solution (\ref{Phit3AB}) clearly indicates that if the clock's energy indeterminacy is large enough, then the state of the two clocks factorizes and the projection postulate applies. However, in this case the energy indeterminacy is too large for Bob, by measuring his measurement apparatus, to gain any useful information about energy changes and hence about Alice's choice to measure her spin or not.

In the other extreme case, if the energy indeterminacy of the clocks is low enough for Alice and Bob to distinguish the cases in which their clocks provided energy, gained energy or ended up with their initial energy, then, as in the spin case of sub-section \ref{simulMeas}, the clocks get entangled. This clock-entanglement remains even once the pointers have been measured. Hence, Bob, by merely measuring his measurement apparatus (pointer plus clock) can gain useful information about whether or not Alice did measure her spin. This confirms that there is signalling, indeed.

In the conclusion we discuss what produces this signalling and how one should interpret and possibly model it. But let's immediately recall that we use a non-local next neighbor interaction Hamiltonian. Hence, at the end of the day, the presence of signalling should not be surprising.

The other main lesson from this exercise is that the projection postulate should not be applied blindly. There are cases in which it leads to wrong predictions. In particular, when measuring additive physical quantities that commute with the global Hamiltonian, hence are conserved quantities, like energy as in our example, one should be careful and better develop a full measurement model rather than merely applying the projection postulate. Somehow, this is not too surprising. Indeed, to measure energy requires time \cite{AMPTime}. But then, during the measurement the system evolves: intuitively, the measured system gets continuously projected, but the components projected at earlier times evolve while some components get projected later, as our model with the quantum clocks illustrate. The latter leads to a reduced state which is, once the clock(s) is(are) traced out, an effectively mixed state.

\section{Conclusion and Outlook}\label{concl}
%===============================
When one spin of an arbitrarily long spin chain is measured, the entire spin chain gets disturbed and its energy modified by the measurement. Assuming next neighbors interactions, this leads to apparent non-physical signalling (section \ref{ABmeasure}) and even to faster than light signalling (appendix \ref{appendixA}). This raises suspicions on energy conservation (see e.g. \cite{SanduEnergy}) - though implausible - and on the brutal application of the projection postulate. Since energy measurements require time \cite{AMPTime}, any explicit measurement model requires the introduction of quantum clocks, as we did in section \ref{meas}. We applied this simple and natural measurement model to our spin chain example, see section \ref{meas3qubits}, concluding that the projection postulate can't be merely applied to all measurements, especially not when conserved additive physical quantities like energy are concerned. This illustrates the WAY theorem \cite{WAYWigner,WAYAraki,WAYYanase,WAYBusch}. Our measurement model, however, confirms the presence of signalling.

This signalling is clearly due to the non-local Hamiltonian that connects next neighbors. This is reminiscent of the Lieb-Robinson theorem which bounds the speed of information flow in such spin systems \cite{LiebRobinson,EisertLR,VerstreateLiebRobinson}. However, this theorem only sets an exponential bound on the amount of information that propagates faster than this speed. Hence, no sharp no-signalling bound can be derived from the Lieb-Robinson result, in particular it has nothing to do with the speed of light. 

\begin{figure}[h]
\includegraphics[width=8.5cm]{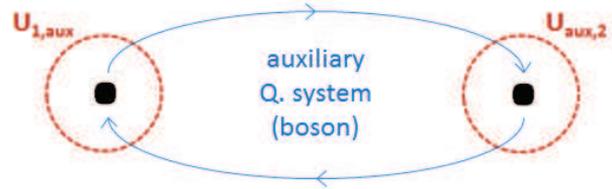}
\vspace{-10pt}
\caption{\it Sketch of a model for non-local Hamiltonians describing the interaction between two quantum systems (black dots), with a flying boson (auxiliary system) and local unitaries $U$.}
\label{FigBoson}
\end{figure}

At the end of our analysis, the explanation of the apparent signalling as entirely due to the use of non-local interaction Hamiltonians is certainly correct, but not fully satisfactory. One may like to model such non-local Hamiltonians with explicit "bosons" that fly at finite speeds between next neighbors whose net result could be summarized by the non-local Hamiltonian. This can be done by merely invoking quantum field theory. However, we believe that a model closer in spirit to quantum information would be very valuable. More precisely, the idea to model an interaction Hamiltonian between two spins would be to add an auxiliary quantum system that flies back-\&-forth between these two spins with local unitary interactions $U$, see Fig. \ref{FigBoson}, such that after a go-\&-return the net result on all eigenstates $\ket{\psi_\varepsilon}$ of the non-local Hamiltonian is a global phase proportional to the Hamiltonian eigenvalues $\varepsilon$:
\beq
U_{1,aux}\otimes U_{aux,2}\ket{\psi_\varepsilon}\otimes\ket{0}=e^{-i\varepsilon\tau/\hbar}\ket{\psi_\varepsilon}\otimes\ket{0}
\eeq
where $\tau$ is the time of a go-\&-return of the auxiliary system and $\ket{0}$ is its initial state. In this way, on time scales longer than $\tau$, the auxiliary system mediates an effective interaction between the two spins described by the Hamiltonian $H=\sum_\varepsilon \varepsilon\ketbra{\psi_\varepsilon}{\psi_\varepsilon}$. The search for such a quantum information inspired model raises interesting questions. Can the auxiliary system merely be a qubit? It seems not. Can it be finite dimensional? or does it need infinite dimensions like the bosons (harmonic oscillators) of quantum field theory?

\small
\section*{Acknowledgment} This work profited from stimulating discussions with Antonio Acin, Nicolas Brunner, Paul Busch, Jens Eisert, Marcus Huber, Sandu Popescu, Valerio Scarani and many participants of the SANDU workshop (May'17) where this work was presented. Financial support by the Swiss NCCR-QSIT is gratefully acknowledged.

\newpage
\appendix

\section{Sharp energy measurements lead to apparent faster than light signalling}\label{appendixA}
%==============================================================================
Section \ref{ABmeasure} illustrated that our spin chain allows for signalling. We termed this signalling non-physical although it is not faster than light, because it assumes post-selecting the cases where Alice and Bob both get result +1. Here we show that there is also faster-than-light signalling.

In this Appendix we use the more compact notations 0 and 1 for the measurements results, instead of $+1$ and $-1$, respectively.

The tables \ref{table1} and \ref{table2} present the energy changes in all possible cases when only Bob measures his spin (Table \ref{table1}) and when both Alice and Bob measure their spins (Table \ref{table2}).

Table I shows that on average the 3-spin system's energy increases by $+\half$, see last column. This corresponds to $+1/3$ if Bob gets spin up ($b=0$), probability 3/4 and +1 if he gets spin down ($b=1$), probability 1/4. If the 3-spin system's energy is measured sharply, then the energy change can only take values 0, 1 and 2 (where 0 means no change). Column 4 indicates the corresponding probabilities.

\begin{table}[h]
	\centering
		\begin{tabular}{c|c|c|c|c|c}
		b&prob(b)&$\Delta E_b$&prob($\Delta E_b$)&$\overline{\Delta E_b}$&$\left\langle\Delta E\right\rangle$ \\
		\hline
		0&3/4&0&3/4&&\\
		&&1&1/6&+1/3&\\
		&&2&1/12&&\\ 
		&&&&&+1/2\\
		1&1/4&0&1/4&&\\
		&&1&1/2&+1&\\
		&&2&1/4&&\\
			
		\end{tabular}
		\caption{The 1st column indicates Bob's measurement result, with the corresponding probability in the 2nd column. The 3rd column indicates the possibly energy increases of the 3-spin system, with the corresponding probabilities (conditioned on Bob's measurement result) in column 4. Column 5 is the mean energy increase conditioned on Bob's result, while the last column is the unconditional energy increase.}\label{table1}
\end{table}
\begin{table}[h]
	\centering
		\begin{tabular}{c|c|c|c|c|c}
		a b&\tiny prob(a,b)&$\Delta E_{ab}$&prob($\Delta E_{ab}$)&$\overline{\Delta E_{ab}}$&$\left\langle\Delta E\right\rangle$\\
		&&\tiny=$\Delta E_{a|b}+\Delta E_{b|a}$&&\tiny=$\overline{\Delta E_{a|b}}+\overline{\Delta E_{b|a}}$&\\
		\hline
		0 0&1/2&0 \tiny($\qe0+0$)&1/2&\\
		&&1&0&+1\tiny($\qe\half+\half$)&\\
		&&2\tiny($\qe1+1$)&1/2&\\ 
		&&&&&\\
		1 0&1/4&0\tiny($\qe0+0$)&1/4&\\
		&&1\tiny($\qe a_1+b_1$)&1/2&+1\tiny($\qe1+0$)&+1\\
		&&2\tiny($\qe a_2+b_2$)&1/4&&\tiny($\qe\half+\half$)\\
		&&&&&\\
		0 1&1/4&0\tiny($\qe0+0$)&1/4&\\
		&&1\tiny($\qe a_3+b_3$)&1/2&+1\tiny($\qe0+1$)&\\
		&&2\tiny($\qe a_4+b_4$)&1/4&\\
			
		\end{tabular}
		\caption{Notation are as in table \ref{table1}, with the addition of which apparatuses provides which part of the energy. For example $+1\qe\half+\half$ indicates that the 3-spin energy increase of one unit is presumably equally provides by both measurement apparatuses (pretty obvious by symmetry in this example). When the split is not obvious, the contributions of each apparatus are indicated by $a_j$ and $b_j$.}\label{table2}
\end{table}

To analyse the data in Table \ref{table2} let's start from the last column. When both Alice and Bob measure, the total energy increase is of +1 unit, averaged over all possible measurement results. By symmetry, it is natural to assume that this +1 energy increase is equally provided by both sides, as shown in the small bracket in the last column of Table \ref{table2}, in perfect agreement with the fact that when only one party measures, it provides an energy increase of +1/2 (last column of Table \ref{table1}).

However, the energy contributions of the two sides depend on which of the three possible measurement results obtains, as indicated on the last but one column of table \ref{table2}. For example, when the results are $00$, then the total energy change is +1 and, again by symmetry both sides should provide half an energy unit. But that is not in agreement with the case where only Bob - or only Alice - performs a measurement, as in such a case he - she - provides only +1/3 energy! This is the conundrum we presented in section \ref{ABmeasure}. Note that this does not correspond to faster than light signalling, as this conundrum assumes that both Alice and Bob get result 0 (spin up), hence they must communicate to each other their result. So, let's assume that in case of result 00, both sides provide energy +1/2, as indicated in the bracket of the last column of Table \ref{table2}.

In the remaining of this appendix \ref{appendixA} we show that the conundrum extends to true faster than light signalling. First, assume that for result 00, both sides provide energy +1/2. Next, as a warm-up, consider the results $a=1$, $b=0$, where symmetry arguments are insufficient. Consistency requires:
\beqa
\overline{\Delta E_{b=0}}&=&\frac{1}{3} \\
&=&\frac{\sum_a prob(a,b=0)\overline{\Delta E_{b=0|a}}}{\sum_a prob(a,b=0)}\\
&=&\frac{\half\cdot\half+\frac{1}{4}\cdot\overline{\Delta E_{b=0|a=1}}}{\half+\frac{1}{4}}
\eeqa
from which one deduces $\overline{\Delta E_{b=0|a=1}}=0$ and $\overline{\Delta E_{a=1|b=0}}=1$, as indicated in the last but one column of table \ref{table2}. By symmetry $\overline{\Delta E_{a=0|b=1}}=0$ and $\overline{\Delta E_{b=1|a=0}}=1$.

Consequently, the mean energy contributions of each apparatuses distributes in a natural way, see last but one column of Table \ref{table2}.

So far so good. But let's now look at the energy changes with a resolution better than the energy unit. Such (relatively) sharp energy measurements require time, but this is not the issue we consider here (anyway, the distance between Alice and Bob could be arbitrarily large - and if needed the number of spins in the chain increased).

Assume first that $a=b=0$ and the 3-spin system remains in the ground state. This happens with a probability $\half\cdot\half=\frac{1}{4}$ as indicated in the first line of table \ref{table2}. In this case, it is ``obvious'' that none of the 2 measurement apparatuses provide any energy, as indicated by $0=0+0$ in table \ref{table2} (the question marks around ``obvious'' is an indication that this is not obvious at all: actually, it is wrong as shown in section \ref{meas3qubits}). The next case that may happen is $a=b=0$ and $\Delta E_{00}=2$, again with a probability $\half\cdot\half=\frac{1}{4}$. By symmetry, these 2 units of energy are provided symmetrically by both apparatuses providing each one unit: $2=1+1$.

Things get more interesting in the case $a=1$, $b=0$ and $\Delta E_{10}=1$ or $\Delta E_{10}=2$. Denote $b_1$ and $b_2$ Bob's measurement apparatus energy contributions for these two cases, respectively, as indicated in Table \ref{table2}. When averaging over Alice's result one should recover the data of table \ref{table1}. 

In the case $\Delta E_b=1$ the conditional probability that $\Delta E_b=1$ should equal to the average over Alice's result:
\beqa
&&prob(\Delta E_b=1|b=0)=\frac{1}{6}\nonumber\\
&=&prob(a=0|b=0)prob(\Delta E_{b|a=0}=1|a=0,b=0)\nonumber\\
&+&prob(a=1|b=0)prob(\Delta E_{b|a=1}=1|a=1,b=0)\nonumber\\
&=&\frac{2}{3}\cdot \half+\frac{1}{3}\frac{\delta(b_1=1)}{2} \ge \frac{1}{3}
\eeqa
which leads to a contradiction: it is impossible that each measuring apparatus provides energy independently of whether the other party performs also a measurement or not! Hence, by measuring sharply the energy change of their measuring apparatus each party should be able to deduce whether the other party also measured her system. This would be true signalling without post-selection. This aspect of the conundrum is also solved by the explicit construction of measurement models that include the measurement apparatus and the clock inside a larger quantum system, as described in section \ref{meas3qubits}.

%=================
% Consider first the case $\Delta E_b=2$. The conditional probability that $\Delta E_b=2$ is 1/12 (table \ref{table1}) which should be equal to the average over Alice's result:
%\beqa
%&&prob(\Delta E_b=2|b=0)=\frac{1}{12}\nonumber\\
%&=&prob(a=0|b=0)prob(\Delta E_b=2|a=0,b=0)\nonumber\\
%&+&prob(a=1|b=0)prob(\Delta E_b=2|a=1,b=0)\nonumber\\
%&=&\frac{2}{3}\cdot 0+\frac{1}{3}\frac{\delta(b_2=2)}{4}
%\eeqa
%which implies $b_2=2$, that is, in case $a=1$, $b=0$ and the 3-spin system increases its energy by 2 units, then all the energy is provided by Bob's apparatus.

\section{Simultaneous measurements on one spin, once the pointers have been measured}\label{appendixB}
%====================================================================================
In this Appendix we analyse the state of the spin and clocks after the two pointers have been measured, starting from the pre-measurement state (\ref{PhitAaB}). Consider first the case where both pointers are found in the initial state $\ket{0}$:
\beqa
\psi_{00}=\big\{ &-&s^3c\ket{\up}e^{-2i\omega\cdot\max(x,y)} \nonumber\\
&-&sc^3\ket{\up}e^{-2i\omega\cdot\min(x,y)} \nonumber\\
&+&s^2c^2\ket{\down} e^{-2i\omega|x-y|} \nonumber\\
&+&c^4\ket{\down}\big\} \cdot\Phi_t(x,y) \label{eqB1}
\eeqa
The square norm of $\psi_{00}$ provides the probability that the two pointers are found in state $\ket{00}$:
\beq
|\psi_{00}|^2=c^2(1-2s^2c^2)+s^2c^4 F
\eeq
where
\beq\label{F}
F=\int\!\!\int dxdy(e^{2i\omega|x-y|}+cc)|\Phi_t(x,y)|^2
\eeq
Similarly for the other possible results on the pointers' measurements:
\beqa
|\psi_{11}|^2&=&s^2(1-2s^2c^2)+s^4c^2 F \\
|\psi_{01}|^2&=&2s^2c^4-s^2c^4 F \\
|\psi_{10}|^2&=&2s^4c^2-s^4c^2 F \label{eqB6}
\eeqa
Note that normalization is preserved, $\sum_{jk}|\psi_{jk}|^2=1$, as it should.

Let's now assume that the clocks are initially in two independent Gaussian states with the same width $\Delta$, i.e.
\beq
\Phi_t(x,y)=\frac{1}{\sqrt{2\pi\Delta^2}}e^{-(x-t+x_i)^2/(2\Delta^2)}e^{-(y-t+y_i)^2/(2\Delta^2)}
\eeq
Then
\beq
F=2\cos(2\omega(y_i-x_i))e^{-4\omega^2\Delta^2}
\eeq

Consider first the case where the energy indeterminacy of the clocks is too large compared to $\omega$ to distinguish energy exchanges, i.e. $\omega\Delta<<1$. In this case $F\approx 2$ and both pointers always give the same result and one recovers the usual projection postulate:
\beqa
&\psi_{00}&=-c\ket{\theta^\perp}\hspace{1cm}\psi_{11}=s\ket{\theta} \label{psi00collapse}\\
&\psi_{01}&=0 \hspace{1.9cm} \psi_{10}=0 \label{psi01collapse}
\eeqa
while the clocks remains disentangled from the spin.

The other limiting case is more interesting. Assume $\omega\Delta>>1$, i.e. the clocks have well enough determined energies to distinguish energy exchanges. In this case $F\approx0$ and the probabilities for the different possible pointers' measurement results read:
\beqa
|\psi_{00}|^2&=&c^2(1-2s^2c^2)\label{psi00}\\
|\psi_{11}|^2&=&s^2(1-2s^2c^2)\\
|\psi_{01}|^2&=&2s^2c^4\\
|\psi_{01}|^2&=&2s^4c^2
\eeqa
If one traces out the clocks, one finds the spin in mixed states. For example:
\beq\label{rho11}
\rho_{11}=s^2(1-2s^2c^2)\left(c^2\ketbra{\up}{\up}+s^2\ketbra{\down}{\down}\right)
\eeq

All of eqs (\ref{psi00}-\ref{rho11}) differ sharply from the predictions of the projection postulate.

\section{Simultaneous measurements of one spin with entangled (squeezed) clocks}\label{appendixC}
%================================================================================
Let's consider the case where the 2 clocks of subsection \ref{simulMeas} are initially entangled in such a way to have a strong correlation in position, i.e. $x\approx y$. The equations (\ref{PhitAB}) and (\ref{eqB1}-\ref{eqB6}) remain valid, except that $F$ has to be re-computed according to the clocks' state.

Assume the initial state of the 2 clocks reads (up to normalization):
\beq
\phi_0(x,y)^2=e^{-(x+y+2x_i)^2/2\Delta_+^2}\cdot e^{-(x-y)^2/2\Delta_-^2}
\eeq
with $\Delta_-<<\Delta_+$. The parameter $F$ defined in (\ref{F}) reads:
\beq
F=2e^{-4\omega^2\Delta_-^2}
\eeq

Since $\Delta_-$ is small, $F\approx 2$ and one recovers eqs (\ref{psi00collapse}) and (\ref{psi01collapse}), i.e. the usual collapse happens and the two pointers always provide the same result.

\section{Global state in case Alice measures before Bob}\label{appendixD}
%=======================================================
We apply the general solution (\ref{Phit3AB}) to the case where Alice measures first and Bob next, such that their clocks don't overlap. Using notations similar to (\ref{PhitAaB}), i.e. $\ket{(x,p)}$ denotes a wave-packet center around mean position $x$ and mean momentum $p$, one gets for the  3-spin chain, together with Alice and Bob's pointers and clocks:
\beqa
&&\ket{\psi_{-1},0_A,0_B,(-x_i,p_a),(-y_i,p_B)}\stackrel{t_{Alice \:measures}}{\longrightarrow} \nonumber\\
&&\ket{\phi_{-1},\frac{3\cdot0+1}{4},0,(t-x_i,p_A),(t-y_i,p_B)} \nonumber\\
&+&\ket{\phi_{+1},\frac{0-1}{4},0,(t-x_i,p_A-2),(t-y_i,p_B)} \nonumber\\
&-&\ket{\phi_{0},\frac{0-1}{4},0,(t-x_i,p_A-1),(t-y_i,p_B)} \\
&&\stackrel{t_{Alice \:and \:Bob \:measure}}{\longrightarrow}\phi_t^{A<B} \nonumber\\
&=&\ket{\phi_{-1},\frac{3\cdot0+1}{4},\frac{3\cdot0+1}{4},(t-x_i,p_A),(t-y_i,p_B)} \nonumber\\
&+&\ket{\phi_{+1},\frac{3\cdot0+1}{4},\frac{0-1}{4},(t-x_i,p_A),(t-y_i,p_B-2)} \nonumber\\
&+&\ket{\phi_{0},\frac{3\cdot0+1}{4},\frac{0-1}{\sqrt{8}},(t-x_i,p_A),(t-y_i,p_B-1)} \nonumber\\
&+&\ket{\phi_{+1},\frac{0-1}{4},\frac{3\cdot0+1}{4},(t-x_i,p_A-2),(t-y_i,p_B)} \nonumber\\
&+&\ket{\phi_{-1},\frac{0-1}{4},\frac{0+-1}{4},(t-x_i,p_A-2),(t-y_i,p_B+2)} \nonumber\\
&-&\ket{\phi_{0},\frac{0-1}{4},\frac{0-1}{\sqrt{8}},(t-x_i,p_A-2),(t-y_i,p_B+1)} \nonumber\\
&+&\ket{\phi_{+1},\frac{0-1}{\sqrt{8}},\frac{0-1}{\sqrt{8}},(t-x_i,p_A-1),(t-y_i,p_B-1)} \nonumber\\
&-&\ket{\phi_{-1},\frac{0-1}{\sqrt{8}},\frac{0-1}{\sqrt{8}},(t-x_i,p_A-1),(t-y_i,p_B+1)} \nonumber\\
&-&\ket{\phi_{0},\frac{0-1}{\sqrt{8}},\frac{0+1}{\sqrt{8}},(t-x_i,p_A-1),(t-y_i,p_B)} \label{Deq}
\eeqa
Each line of (\ref{Deq}) shows all the possible energy exchanges. For example the second line corresponds to the case where the 3-spin chain gained 2 energy units, provided by Bob's clock; the last bit one line corresponds to the case where the 3-spin chain ends in the initial state $\phi_{-1}$ but exchanged energy with both clocks such that eventually Alice's clock lost one energy unit and Bob's clock gained this energy unit.


\begin{thebibliography}{99}
\bibitem{AMPTime} Y. Aharonov, S. Massar, S. Popescu,  Phys. Rev. A {\bf 66}, 052107 (2002).
\bibitem{LiebRobinson} E. H. Lieb and D. W. Robinson, Commun.Math. Phys. 28, 251 (1972). For a recent review see, e.g. \cite{EisertLR}
\bibitem{EisertLR} M. Kliesch, Ch. Gogolin and J. Eisert, in {\it Many-Electron Approaches in Physics, Chemistry and Mathematics},  L. D. Site and V. Bach (eds.), (Springer: Mathematical Physics Studies, 2014), pp. 301-318
\bibitem{VerstreateLiebRobinson} S. Bravyi, M.B. Hasting and F. Verstreate, Phys. Rev. Lett. {\bf 97}, 050401 (2006).
\bibitem{WAYWigner} E. Wigner Z. Phys. {\bf133}, 101 (1952)
\bibitem{WAYAraki} H. Araki and M.M. Yanase, Phys. Rev. {\bf120}, 622–6 (1960)
\bibitem{WAYYanase} M.M. Yanase, Phys. Rev. {\bf123}, 666 (1961)
\bibitem{WAYBusch} L. Loveridge and P. Busch Eur. Phys. J. D {\bf62}, 297 (2011)
\bibitem{BuschQTM} P. Busch, P.J. Lathi and P. Mittelsteadt, {\it The Quantum Theory of Measurement}, Springer 1996.
\bibitem{GisinQchance14} N. Gisin, {\it Quantum Chance, nonlocality, teleportation and other quantum marvels}, Springer, 2014.
\bibitem{QMVN} J. Von Neumann, {\it Mathematical Foundations of Quantum Mechanics}, Princeton University Press, Princeton, New Jersey, 1955.
\bibitem{AharonovBohmTime} Y. Aharonov and D. Bohm, Phys. Rev. {\bf 122}, 1649 (1961).
\bibitem{weakValue} Y. Aharonov, D. Z. Albert, and L. Vaidman, Phys. Rev. Lett. {\bf60}, 1351 (1988).
\bibitem{AharonovQrefframe} Y. Aharonov and T. Kaufherr, Phys. Rev. D {\bf 30}, 368 (1984).
\bibitem{AharonovArrivalTime} Y. Aharonov, J. Oppenheim, S. Popescu, B. Reznik and W.G. Unruh, Phys. Rev. A {\bf 57}, 4130 (1998).
\bibitem{SanduEnergy} Y. Aharonov, S. Popescu and D. Rohrlich, arXiv:1609.05041 

\end{thebibliography}
\end{document}